 \documentclass[10pt,preprint2]{aastex}

\usepackage{graphicx}
\usepackage{amsmath}

\def \apj {ApJ}

\def \apjl {ApJ}
\def \solphys {Solar Phys.}

\def \aap {A\&A}

\newcommand{\citeN}[1]{\citeauthor{#1} (\citeyear{#1})}
\newcommand{\citeNP}[1]{\citeauthor{#1} \citeyear{#1}}

\shortauthors{Socas-Navarro and Manso Sainz}
\shorttitle{Shocks in the Quiet Sun}

\begin{document}

\title{Shocks in the Quiet Solar Photosphere: A Rather Common Occurrence}

\author{H. Socas-Navarro and R. Manso Sainz}
   	\affil{High Altitude Observatory, NCAR\thanks{The National Center
	for Atmospheric Research (NCAR) is sponsored by the National Science
	Foundation.}, 3450 Mitchell Lane, Boulder, CO 80307-3000, USA}
	\email{navarro@ucar.edu}

\date{}%

\begin{abstract}
We present observations of the quiet solar photosphere in the \ion{Fe}{1}
lines at 6302 \AA \, where at least four
different spatial locations exhibit upwards-directed supersonic flows. These
upflows can 
only be detected in the circular polarization profiles as a double-peaked 
structure in the blue lobe of both \ion{Fe}{1} lines. We have detected cases
of either magnetic polarity in the data. The polarization
signals associated with the upflows are very weak, which is probably why they
had not been seen before in this type of observations. We propose that the
observed flows are the signature of aborted convective collapse, similar to
the case reported by \citeN{BRRHC+01}. Our data indicates that this
phenomenon occurs frequently in the quiet Sun, which means that many magnetic
elements (although the fraction is still unknown) are destroyed even before
they are formed completely. The spectral signatures of supersonic upflows
reported here are probably present in most spectro-polarimetric observations
of sufficient signal-to-noise and spatial resolution.
\end{abstract}


\section{Introduction}
\label{sec:intro}

The term {\it quiet Sun} is commonly used to refer to the solar atmosphere
away from sunspots, faculae and other large-scale concentrations of magnetic 
flux. The adjective ``quiet'', however, seems now less of a synonym of
``unintersting'' 
than it did only a 
few years ago. This is due to both recent numerical simulations
(e.g., \citeNP{EC01}; \citeNP{CSN+04}) and new spectro-polarimetric
observations of high sensitivity and resolution (e.g., \citeNP{SAL00};
\citeNP{SNSA02}; \citeNP{KCS+03}; \citeNP{DCKSA03}). From the point of view
of the magnetic energy balance it is worth mentioning that, according to
these works, there is probably more (unsigned) magnetic flux in the quiet
Sun than in all the active regions during solar maximum. The small-scale
structure of the magnetic elements is also a hot topic of debate
(e.g., \citeNP{SNSA03}). 

The present work adds to this more exciting view of the quiet
photosphere by reporting on the detection of supersonic upflows, which seem
to be a relatively frequent event. We argue that these flows are likely 
the last stages of the process observed by \citeN{BRRHC+01} (hereafter BRCKR)
in the 
infrared. According to their scenario, the upflows are the signature of an
aborted convective collapse 
and subsequent disintegration of the magnetic structure. Our
observations support this picture to some extent and remove any
caveats that might have existed on their observation (e.g., structures
moving in or out of the slit). 

\section{Observations}
\label{sec:obs}

The observations used in this paper were taken with the Advanced Stokes
Polarimeter (ASP, \citeNP{ELT+92}) and first analyzed by \citeN{L96}. The
strong upflows that we report here have gone unnoticed until now because
their associated polarization signals are very weak. The amplitudes of
Stokes~$V$ are at most of the order of 5$\times$10$^{-3}$, and even
weaker in some of the interesting points (at the 10$^{-3}$ level). This value
is close to the noise level of the observations (5$\times$10$^{-4}$), but
fortunately it is high enough that it allows for a few evident detections.

\begin{figure*}
\epsscale{1.6}
\plotone{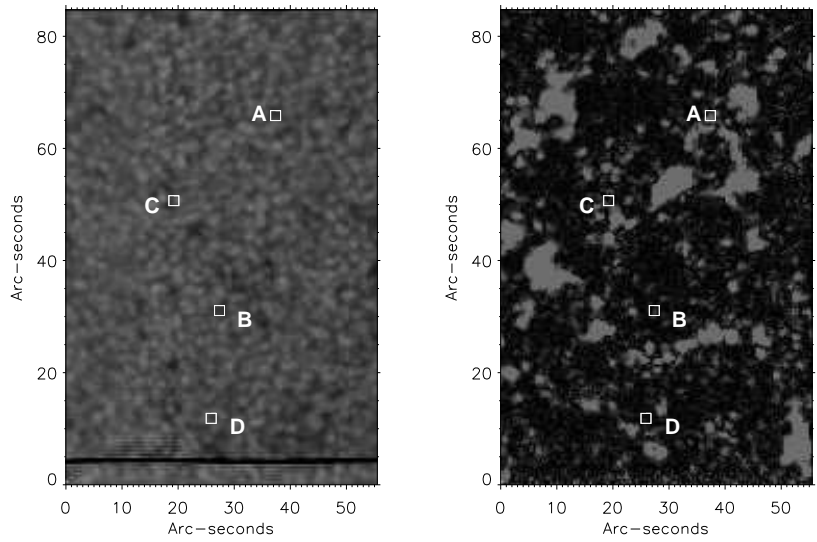}
\caption{
\label{fig:maps}
Maps of the observed region showing the location of the supersonic
upflows. Left: Photospheric continuum. Right: Degree of circular polarization
(V/I integrated over a 130~m\AA \, bandwidth) saturated at 0.003. White
squares mark the position of interesting profiles (see text).
}
\end{figure*}

Figure~\ref{fig:maps} shows a map of the continuum intensity in the observed
region (left) and the degree of circular polarization (right). The four
spatial locations of regions where we have found clear 
detections are marked with A, B, C and D, respectively. We have found
some other pixels where the profiles seem to exhibit a similar structure, but
the signal-to-noise ratio is not good enough to be sure. Three of the
detections have the same polarity (regions A, B and C), whereas the forth
one (D) has the opposite. 

\begin{figure*}
\plotone{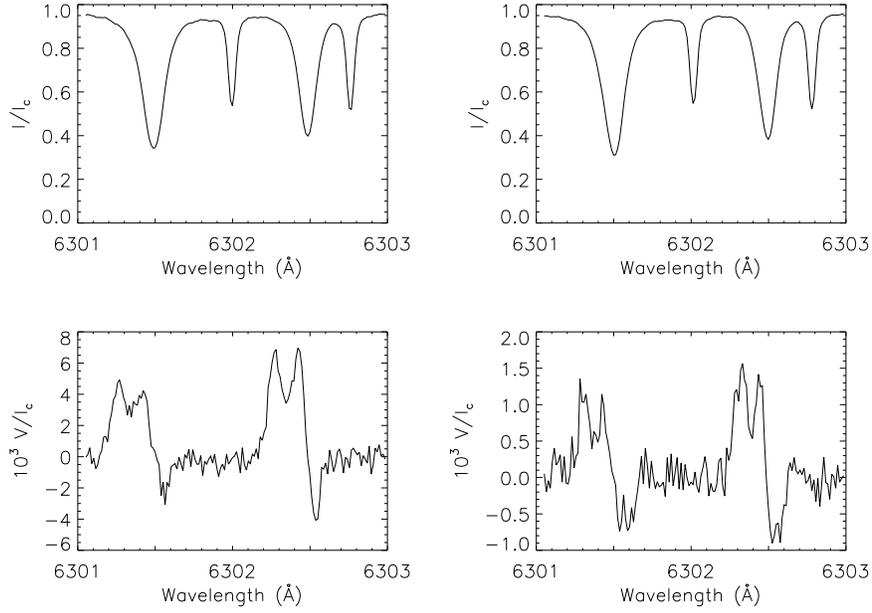}
\caption{
\label{fig:profs}
Individual Stokes~$I$ (upper panels) and~$V$ (lower panels) spectra of the
\ion{Fe}{1} lines at 6301.5 and 6302.5 \AA . The narrow 
lines at 6302.0 and 6302.8 are telluric. The profiles correspond to
regions~A (left panels) and~B (right panels), and
have been normalized to the local continuum intensity $I_c$.
}
\end{figure*}

The Stokes~$I$ and~$V$ profiles corresponding to the best two detections
are plotted in Fig~\ref{fig:profs}. The intensity profiles do not show any
distinctive features. However, the Stokes~$V$ spectra show a double-peaked
structure in the blue side of the line. Observational studies of Stokes~$V$
asymmetries in the quiet Sun show that the red lobe is typically broader and
shallower than the blue one (e.g., \citeNP{SAL00}), which explains why the
double peak is not 
visible in the red lobe of the lines. The zero-crossing wavelength of the
overall profile is practically at the same position as the intensity
minimum. The difference between the two blue peaks in both lines 
corresponds approximately to 7.5~km~s$^{-1}$. 

These two profiles were the first that we identified in the map by visual
inspection. We developed an automated search algorithm to try to find other
similar occurrences in the map. The algorithm looks for extended blue wings
in both \ion{Fe}{1} lines. The candidates detected in this manner have
been identified in Fig~\ref{fig:maps}.

\begin{figure*}
\plotone{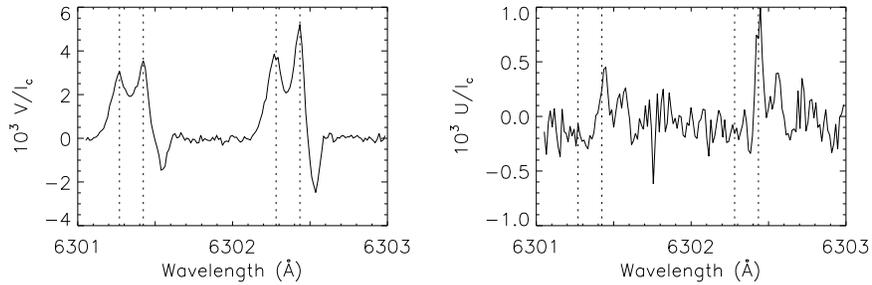}
\caption{
\label{fig:avprofs}
Spatial average of Stokes~$V$ (left) and~$U$ (right) profiles in region~A. The 
average has been computed over a 4$\times$4 pixel box. The vertical
lines mark the positions of the Stokes~$V$ blue peaks.
}
\end{figure*}

The upflow in region~A spans an area large enough that we can improve the
signal-to-noise ratio considerably by binning the profiles spatially. Using
a 4$\times$4 box average we were able to detect some linear polarization
signal in Stokes~$U$ (Stokes~$Q$ is still within the noise after
binning). Figure~\ref{fig:avprofs} shows the binned Stokes~$V$ and~$U$. It is
interesting to note that the upflowing component does not have any 
linear polarization associated to it in either \ion{Fe}{1}
line. 

We also looked for profiles similar to these in other high-resolution quiet
Sun observations. For this purpose we analyzed the three maps observed by
\citeN{LSN04} with the Diffraction-Limited Stokes Polarimeter
(\citeNP{SEL+03}). Although those maps have very good spatial resolution
($\simeq$0.6''), we could not find any obvious occurrence. We believe the
reason is the lower signal-to-noise ratio in the DLSP data (about a factor 3
worse).

\section{Interpretation}
\label{sec:interpret}

The most natural explanation for the double-peaked structure in Stokes~$V$ is
the presence
of two atmospheric components with very different velocities along the
line of sight (i.e., in the vertical direction, since these observations were
taken at disk center). 
Inspection of the Stokes~$V$ spectra along the slit shows that the red
component of the double peak is nearly at rest with respect to the other
magnetic signals. We shall refer to this as the ``normal'' component. The
blue component is shifted in wavelengths by at least 157~m\AA \, (the
separation between the peaks), or approximately 7.5~km~s$^{-1}$. This will be
referred to as the ``upflowing'' component.

For a more detailed study, we carried out inversions of the average profiles
in region~A. We used a modified version of the LTE code LILIA
(\citeNP{SN01a}; see also \citeNP{RCdTI92}) that considers two magnetic
atmospheres embedded in an unmagnetized background. In addition to the usual
magnetic filling factor $\alpha$, we now have the individual filling factors
of the two magnetic components. The non-magnetic component is assumed to
produce an average unpolarized quiet Sun profile. The two magnetic components
are allowed to have different temperature, density, line of sight velocity,
magnetic field vector (strength and orientation) and microturbulence. These
parameters are recovered independently by the inversion code as a function of
height. 

\begin{figure*}
\plotone{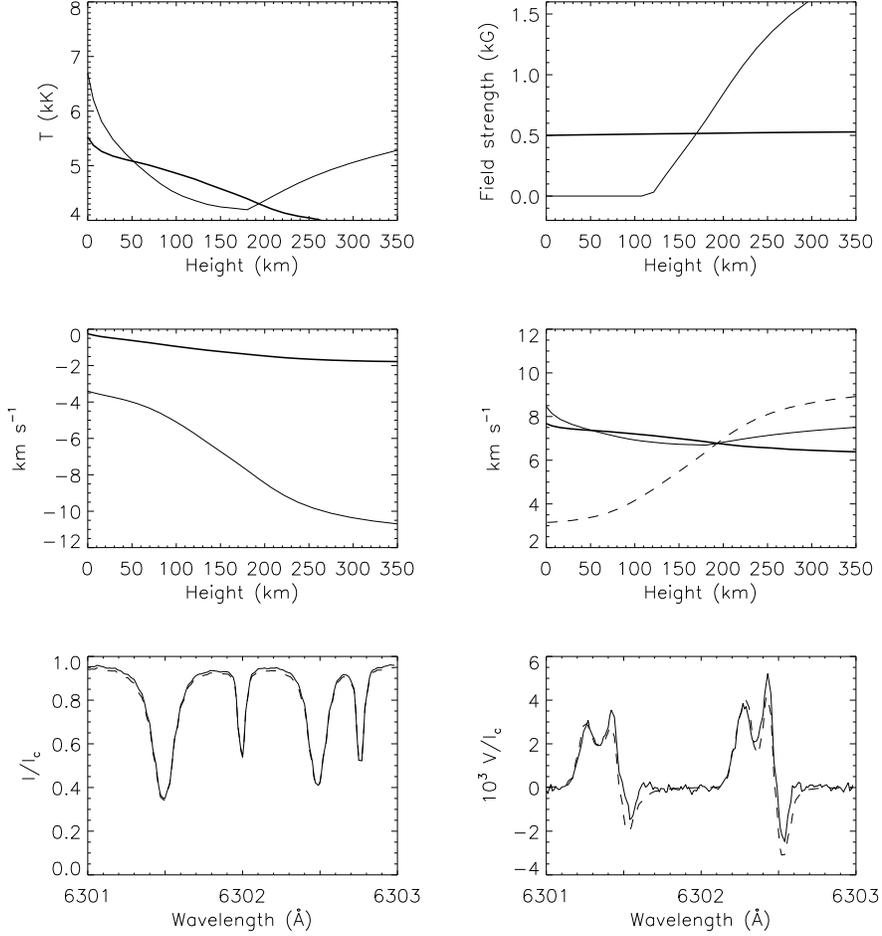}
\caption{
\label{fig:inv}
Model atmospheres (top and middle panels) and fits (bottom panels) obtained
from a 2-component inversion. 
The thick (thin) line represents physical variables
in the normal (upflowing) component. The middle panels show the line of sight
velocities (left) and the sound speed (right) in each atmospheric
component. The dashed line in the middle right panel is the relative velocity
between the two components. The bottom panels show the observed
Stokes~$I$ and~$V$ profiles (solid) and the fits provided by the code
(dashed). The magnetic filling factor retrieved is $\alpha=0.95$. The
relative occupation of the normal and upflowing components are 0.7 and 0.3,
respectively. 
}
\end{figure*}

The models obtained from the inversion, as well as the fits to Stokes~$I$
and~$V$, are presented in Fig~\ref{fig:inv}. While the normal component
exhibits a rather smooth behavior with more or less constant properties, the
upflowing component shows some interesting properties. The temperature stops
decreasing with height and even increases considerably above $z=200$~km,
approximately. The magnetic field, which has a more or less constant value of
500~G in the normal component, has a sharp discontinuity going from zero to
1~kG between $z=150$ and $z=200$~km. Finally, the line of sight velocity
goes from a moderate upflow the bottom of the photosphere to reach values of
$\simeq$10~km~s$^{-1}$ at $z=200$~km. Remarkably, the relative velocity
between the two atmospheres turns from subsonic to supersonic precisely at
$z=200$~km. 

All the properties mentioned above indicate the presence
of a shock front propagating upwards in the photosphere. The coincidence
between the  
three curves in the middle right panel of Fig~\ref{fig:inv} (within a few km)
is remarkable. This means that the point where the upflowing atmosphere is
heated by the shock and its temperature starts to increase is also the point
where the upflow reaches the sound speed in the medium. Since the three curves
are derived independently of each other, this is a good consistency test for
the reliability of the inversion. The strong gradients in the magnetic field
and velicity (present only in the upflowing component) are also consistent
with the shock scenario. 

%

We find that the observations described here bear a striking resemblance to
the infrared ones of BRCKR near 1.6~$\mu$m. Their analysis
of fixed-slit time-series of a quiet Sun region lead them to conclude that
they had observed an instance of aborted convective collapse. After an
initial stage of field concentration with moderate downflows, the magnetic
structure quickly disintegrated presumably destroyed by an
upwards-propagating shockwave. 

We propose that the double-peaked blue lobes of the profiles that we have
detected are the visible counterparts of the last stages of evolution 
in the BRCKR observations. In principle there is no reason why the 
process they observed would be more easily detectable in the 
infrared than in the visible. One might argue that the Doppler shift
increases linearly with 
wavelength, but on the other hand so does the thermal width of spectral
lines. Therefore, the relative velocity needed to produce a double-peaked
Stokes~$V$ lobe would be approximately the same in the visible and the
infrared, especially considering that the spectral lines observed are from
the same ion.

The area fraction occupied by pixels harboring shocks in our map
is $f=$7$\times$10$^{-4}$. Let us assume, for the sake of argument, that
the convective collapse 
observed by BRCKR is ``typical'' and take the duration of the
entire process to be 15~minutes and the upflowing phase 
3~minutes (as reported by them). With these parameters we can estimate that
the probability of these authors to observe an aborted convective collapse in
their 60~minute time-series of a 30$\times$0.5~arc-seconds area is
approximately 0.45. This figure supports the idea that we have observed the
same phenomenon.

The absence of linear polarization in the upflowing component
(Fig~\ref{fig:avprofs}), while the quiet component exhibits some signal,
is surprising. In principle there are two possible explanations for
this. The most obvious would be that the upflow disrupts the initial
organization of the field in the horizontal direction. A second possibility
is that the field may be weaker in the upflowing component, reducing the
linear polarization below the detection threshold. Notice that Stokes~$V$ can
still have a similar amplitude in both components because it would be
compensated by a hypothetical temperature (and thus also source function)
increase in the upflow. This 
explanation is plausible because, while Stokes~$V$ has a first-order
dependence on the field strength, the dependence of Stokes~$Q$ and~$U$ is of
second order. The response to temperature (through the source function) is
the same in all Stokes parameters. Since temperature and field strength
affect linear and circular polarization differently, it is possible to find a
suitable combination that preserves one and destroys the other.

\section{Conclusions}
\label{sec:conc}

The observations presented in this paper reveal supersonic upflows in the
quiet solar photosphere and indicate that this is a rather common
occurrence. The upflows are probably the signature of an exploding magnetic
element after an aborted convective collapse proposed by
BRCKR. It is important to point out that we do not observe the
flux destruction directly. However, the assumption that we are observing the
same process seems appropriate based on the following facts:
\begin{itemize}
\item The only observations reported in the literature with upflows inside
  photospheric magnetic elements is the aborted convective collapse of BRCKR. 
\item The magnitude of the upflows is the same, very close to the sound speed
  of the photosphere at the heights where the lines form. This is what
  one would expect from a shockwave propagating inside the magnetic element.
\item If one assumes that we are observing the same process and that their
  collapse time scales are typical, it is possible
  to estimate the probability that BRCKR observed an aborted
  convective collapse in their time-series. We find that this probability is
  of the order of 1, which would be an unlikely coincidence if our assumption
  were incorrect.
\end{itemize}

From a 2-component inversion of the double-peaked profiles we obtain a model
atmosphere that is consistent with a shockwave propagating in one of the
components. The inversion even retrieves a zero field below the front shock,
which would also be in agreement with the destruction of the magnetic element
reported by BRCKR. 

Interestingly, we only observe instances in which the
two atmospheric components have similar Stokes~$V$ amplitudes. A possible
explanation to this could be the following. The upflowing component cannot
have a larger amplitude because, if it is indeed an exploding magnetic
element, the field must be weaker in this component. If, on the other hand,
it had a significantly lower amplitude then its Stokes~$V$ profile would be
buried in the blue lobe of the quiet component and we would not be able to
see it. Therefore, it is likely that we are only observing a small fraction
of all exploding elements. 

Because of the selection effect mentioned above and the fact that we are
close to the sensitivity limit of the data, the frequency of
occurrence reported here should be taken as a (perhaps severley
underestimated) lower limit.

Finally, we find that the spatial extent of the upflows are quite puzzling,
especially region~A. The upflows span here a significantly
large area of more than 2 by 2 arc-seconds. Considering that the magnetic
filling 
factor is very small in this area ($\sim$10\%), this is most certainly not a
single magnetic structure disintegrating. However, if this area is composed
of a collection of small magnetic elements (as is generally thought that
network elements are), then what physical mechanism would syncrhonize the
exploding elements as far as 2 arc-seconds apart?

It is important to continue investigating these upflows in more detail, given
the potential implications of this work on the generation and destruction of
small-scale magnetic elements in the quiet Sun. As mentioned earlier, these
fields may be carrying most of the solar magnetic flux. It is fundamental to 
determine whether 
convective collapse is an efficient mechanism for concentrating field and
these aborted instances constitute only a small fraction of the cases, or if
on the contrary most (or perhaps even all) of the collapsing elements
are ultimately doomed to disintegrate sending shockwaves into the
photosphere. The amount of energy deposited in the photosphere, and
also in the upper atmosphere, by the shocks might be important if a large
number of these events are taking place continuously in the quiet Sun.
Clearly, further study is required with both new simulations and 
observations to shed some light into these issues.

\acknowledgments
The authors are grateful to B.W. Lites for providing us with the ASP
observations.
\clearpage

\begin{thebibliography}{13}
\expandafter\ifx\csname natexlab\endcsname\relax\def\natexlab#1{#1}\fi

\bibitem[{{Bellot Rubio} {et~al.}(2001){Bellot Rubio}, {Rodr{\'{\i}}guez
  Hidalgo}, {Collados}, {Khomenko}, \& {Ruiz Cobo}}]{BRRHC+01}
{Bellot Rubio}, L.~R., {Rodr{\'{\i}}guez Hidalgo}, I., {Collados}, M.,
  {Khomenko}, E., \& {Ruiz Cobo}, B. 2001, \apj, 560, 1010

\bibitem[{{Carlsson} {et~al.}(2004){Carlsson}, {Stein}, {Nordlund}, \&
  {Scharmer}}]{CSN+04}
{Carlsson}, M., {Stein}, R.~F., {Nordlund}, {\AA}., \& {Scharmer}, G.~B. 2004,
  \apjl, 610, L137

\bibitem[{{Dom{\'{\i}}nguez Cerde{\~ n}a} {et~al.}(2003){Dom{\'{\i}}nguez
  Cerde{\~ n}a}, {Kneer}, \& {S{\' a}nchez Almeida}}]{DCKSA03}
{Dom{\'{\i}}nguez Cerde{\~ n}a}, I., {Kneer}, F., \& {S{\' a}nchez Almeida}, J.
  2003, \apjl, 582, L55

\bibitem[{{Elmore} {et~al.}(1992){Elmore}, {Lites}, {Tomczyk}, {Skumanich},
  {Dunn}, , {Schuenke}, {Streander}, {Leach}, {Chambellan}, {Hull}, \&
  {Lacey}}]{ELT+92}
{Elmore}, D.~F., {Lites}, B.~W., {Tomczyk}, S., {Skumanich}, A., {Dunn}, R.~B.,
  , {Schuenke}, J.~A., {Streander}, K.~V., {Leach}, T.~W., {Chambellan}, C.~W.,
  {Hull}, \& {Lacey}, L.~B. 1992, in Proc SPIE, Vol. 1746, 22

\bibitem[{{Emonet} \& {Cattaneo}(2001)}]{EC01}
{Emonet}, T., \& {Cattaneo}, F. 2001, \apjl, 560, L197

\bibitem[{{Khomenko} {et~al.}(2003){Khomenko}, {Collados}, {Solanki}, {Lagg},
  \& {Trujillo Bueno}}]{KCS+03}
{Khomenko}, E.~V., {Collados}, M., {Solanki}, S.~K., {Lagg}, A., \& {Trujillo
  Bueno}, J. 2003, \aap, 408, 1115

\bibitem[{{Lites}(1996)}]{L96}
{Lites}, B.~W. 1996, \solphys, 163, 223

\bibitem[{{Lites} \& {Socas-Navarro}(2004)}]{LSN04}
{Lites}, B.~W., \& {Socas-Navarro}, H. 2004, \apj, 613, 600

\bibitem[{{Ruiz Cobo} \& {del Toro Iniesta}(1992)}]{RCdTI92}
{Ruiz Cobo}, B., \& {del Toro Iniesta}, J.~C. 1992, \apj, 398, 375

\bibitem[{{S{\'a}nchez Almeida} \& {Lites}(2000)}]{SAL00}
{S{\'a}nchez Almeida}, J., \& {Lites}, B.~W. 2000, \apj, 532, 1215

\bibitem[{{Sankarasubramanian} {et~al.}(2003){Sankarasubramanian}, {Elmore},
  {Lites}, {Sigwarth}, {Rimmele}, {Hegwer}, {Gregory}, {Streander}, {Wilkins},
  {Richards}, \& {Berst}}]{SEL+03}
{Sankarasubramanian}, K., {Elmore}, D.~F., {Lites}, B.~W., {Sigwarth}, M.,
  {Rimmele}, T.~R., {Hegwer}, S.~L., {Gregory}, S., {Streander}, K.~V.,
  {Wilkins}, L.~M., {Richards}, K., \& {Berst}, C. 2003, in Polarimetry in
  Astronomy. Edited by Silvano Fineschi . Proceedings of the SPIE, Volume 4843,
  pp. 414-424 (2003)., 414--424

\bibitem[{{Socas-Navarro}(2001)}]{SN01a}
{Socas-Navarro}, H. 2001, in ASP Conf. Ser. 236: Advanced Solar Polarimetry --
  Theory, Observation, and Instrumentation, 487

\bibitem[{{Socas-Navarro} \& {S{\' a}nchez Almeida}(2002)}]{SNSA02}
{Socas-Navarro}, H., \& {S{\' a}nchez Almeida}, J. 2002, \apj, 565, 1323

\bibitem[{{Socas-Navarro} \& {S{\' a}nchez Almeida}(2003)}]{SNSA03}
{Socas-Navarro}, H., \& {S{\' a}nchez Almeida}, J. 2003, \apj, 593, 581

\end{thebibliography}

\clearpage

\clearpage


\clearpage

%

\end{document}